\documentstyle[art10,hip-100a,epsf]{article}
\begin{document}
\vspace*{-4.5cm}
{\hfill ORNL-CTP-96-09 (hep-ph/9607316)}
\vskip 3cm
\vskip 0.3cm
\cimo \setcounter{page}{1} \thispagestyle{empty} \hskip -15pt  
\noindent
{\Large\bf
Effects of Final-State Interaction and Screening on Strange- and
Heavy-Quark Production 
}\\[3mm] 
\def\rightmark{Final-State Interaction and Screening 
}\def\leftmark{Cheuk-Yin Wong}
\hspace*{6.327mm}\begin{minipage}[t]{12.0213cm}{\large\lineskip .75em
Cheuk-Yin Wong$^1$ and Lali Chatterjee$^{1,2,3}$
}\\[2.812mm] 
\hspace*{-8pt}$^1$ Oak Ridge National Laboratory, Oak Ridge, TN 37831,
U.S.A. \\[0.2ex]
\hspace*{-8pt}$^2$ University of Tennessee, Knoxville, TN 37996,
U.S.A. \\[0.2ex]
\hspace*{-8pt}$^3$ Jadavpur University, Calcutta 700032, India
\\[4.218mm]{\it
Received nn Month Year (to be given by the editors)
}\\[5.624mm]\noindent
{\bf Abstract.}  Final-state interaction and screening have a great
influence on $q \bar q$ production cross sections, which are
important quantities in many problems in quark-gluon plasma physics.
They lead to an enhancement of the cross section for a $q \bar q$
color-singlet state and a suppression for a color-octet state.  The
effects are large near the production threshold.  The presence of
screening gives rise to resonances for $q\bar q$ production just above the
threshold at specific plasma temperatures. These resonances,
especially $c\bar c$ and $b \bar b$ resonances, may be
utilized to search for the quark-gluon plasma by studying the
temperature dependence of heavy-quark pair production just above the
threshold.
\end{minipage}

\section{Introduction}
 
The cross sections for the production of $q \bar q$ pairs are
important quantities in many problems in high-energy heavy-ion
collisions.  For example, the rate of approach to chemical equilibrium
depends on the $s \bar s$ production cross section \cite{Raf96,Won94},
and the charm signal and background depend on the $c \bar c$
production cross section \cite{Mos95,Gav96}.

In the lowest-order evaluation of $q \bar q$ production cross
sections, the quark and the antiquark are described by plane waves,
and their final-state interaction is not included.  The quark and the
antiquark however interact with each other through a color-Coulomb
interaction, which has a great influence on the probability of $q\bar
q$ production. The effects of final-state interactions can be
approximately included in terms of a multiplicative $K$-factor
\cite{Fie89}-\cite{Cha95b}.  While one can use the lowest-order
\cite{Raf96} or the next-to-leading order results \cite{Gav96} for
$q\bar q$ production, there are situations in which a perturbation
expansion involving only the lowest two orders may not be sufficient,
as in the case of large coupling constants and energies near the
production threshold.  A nonperturbative correction, based on the use
of the distorted wave function in the presence of the color-Coulomb
potential, can provide non-perturbative corrections to the cross
sections \cite{Gus88}-\cite{Cha95b}. Furthermore, in the quark-gluon
plasma, the color-Coulomb interaction between a quark and an antiquark
is screened to become a color-Yukawa interaction, and the screening
needs\break
\vskip 0.3cm
\hrule Invited talk presented at Strangeness'96, Budapest, Hungary,
May 15-17, 1996.  To be published in {\it Heavy-Ion Physics}.
\newpage
%
%
\noindent
to be taken into account when we study reactions in
the quark-gluon plasma \cite{Won96c}.

\section{The Importance of Final-State Interactions} 

The importance of final-state interaction can be assessed by studying
the process $e^+ \!+ e^- \!\!\rightarrow$hadrons, which can be
considered to go initially through the reaction $e^+ + e^- \rightarrow
q + \bar q ,$ where the produced $q\bar q$ pair hadronizes
subsequently to produce hadrons.  The total cross section is
\begin{eqnarray}
\label{eq:sigh}
\sigma(e^+ e^- \rightarrow {\rm hadrons}) = N_c \sum_f \biggl ( {e_f
\over e} \biggr )^2 \sigma_f(s),
\end{eqnarray}
\vskip -0.3cm
\noindent
where $\sigma_f$ in lowest-order QCD is 
\begin{eqnarray}
\label{eq:sigf}
\sigma_f(s)= {4\pi \over s} \alpha^2 \sqrt{ 1 - {4 m_f^2 \over s}}
\biggl (1 + {2m_f^2 \over s} \biggr ) \theta \biggl ({ 1 - {4 m_f^2
\over s}} \biggr ) .
\end{eqnarray}
Here, $N_c=3$ is the number of colors, $e_f$ and $m_f$ are the
electric charge and rest mass of the quark $q_f$ with flavor $f$, and
$\alpha$ is the fine-structure constant.  A convenient quantity to use
to compare with experiment is the hadronic to muonic cross section
ratio
\vskip -0.4cm
\begin{eqnarray}
R = {\sigma ( e^+ e^- \rightarrow {\rm hadrons}) \over \sigma(e^+
e^- \rightarrow \mu^+ \mu^-)}. 
\end{eqnarray}
\vspace*{-1.5cm}
\hskip -1.0cm
\epsfxsize=350pt
\includegraphics{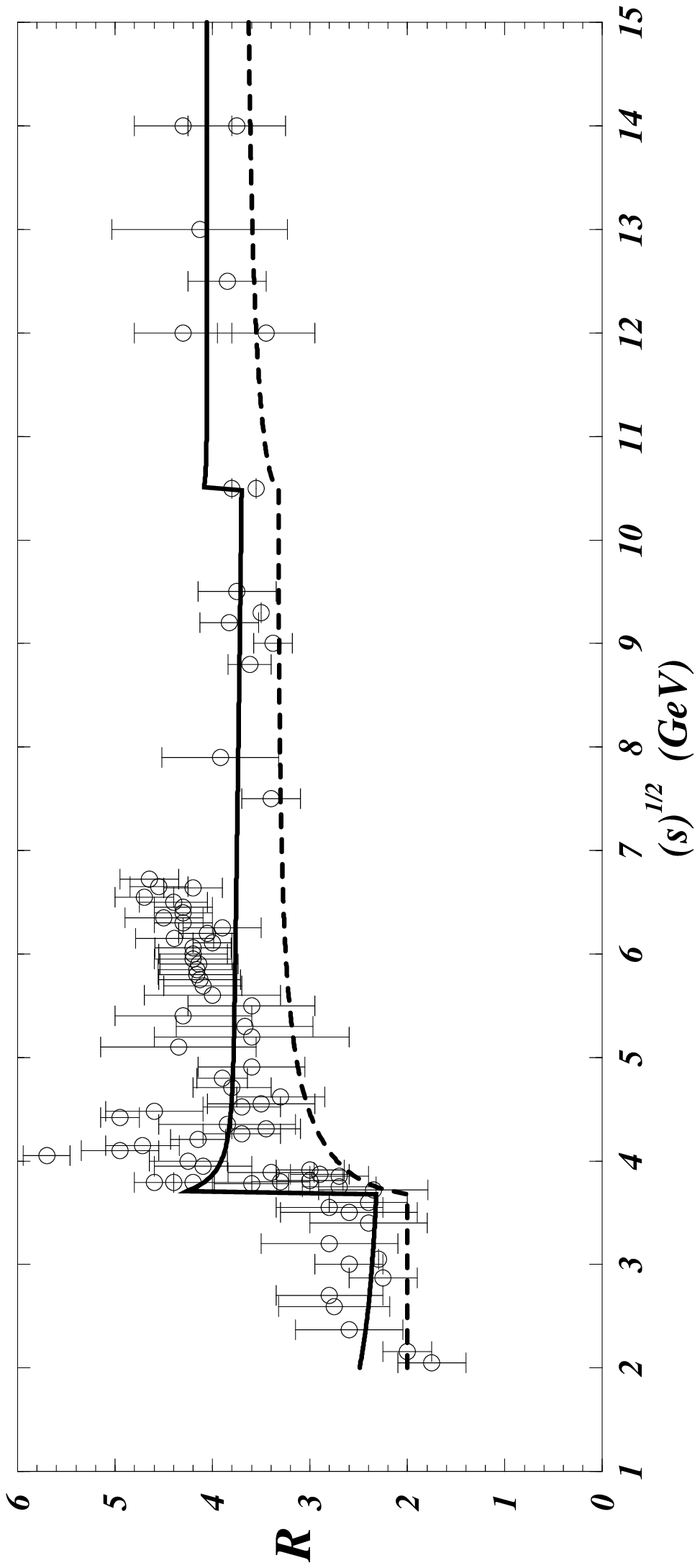}
\vskip 6.4cm
\begin{minipage}[t]{11cm}
\noindent \bf Fig.1.  \rm
The ratio  $
R = {\sigma ( e^+ e^- \rightarrow {\rm hadrons}) / \sigma(e^+
e^- \rightarrow \mu^+ \mu^-)}$ as a function of $\sqrt{s}$.
The dashed curve is the lowest-order result.  The solid curve is
obtained by using the $K$-factor of Eq.\ (\ref{eq:kfac}).
\end{minipage}
\vskip 4truemm
\noindent
The ratios $R$ calculated with Eqs.\ (\ref{eq:sigh}) and
(\ref{eq:sigf}) are shown as the dashed curve in Fig.\ 1 and are
compared with experimental data. The departure of the experimental
data from the lowest-order results of (\ref{eq:sigh})-(\ref{eq:sigf})
can be assessed by defining an empirical multiplicative $K$-factor for
$q\bar q$ production:
\begin{eqnarray}
\label{eq:sighc}
\sigma(e^+ e^- \rightarrow {\rm hadrons}) = N_c \sum_f \biggl ( {e_f
\over e} \biggr )^2  K_f(s) \sigma_f(s).
\end{eqnarray}

A comparison of Eq.\ (\ref{eq:sighc}) and the experimental $R$ ratios
indicates that the empirical values of $K_c(s)$ for $c\bar c$
production are near unity for $\sqrt{s} >> 2m_c$, as predicted by
perturbative QCD \cite{Fie89}.  However, $K_c(s)$ is much greater than
unity near the threshold.  The large deviation from unity arises
predominantly from the final-state interaction between $q$ and $\bar
q$.  The magnitude of the deviation indicates that close to the
threshold a perturbative treatment, including only the leading order
and the next-to-leading order, is inadequate.  Similar effects of
initial- and final-state interactions are known in many other areas of
physics \cite{Gam28,Gyu81}.  The lowest-order QCD results need to be
corrected by a non-perturbative $K$-factor to bring the theoretical
results into approximate agreement with the observed values.

\section{K-factor for Color-Coulomb Interactions} 

To study the corrections to the lowest-order cross sections for $q
\bar q$ production, we consider the case of a small relative velocity
between $q$ and $\bar q$.  The color-Coulomb interaction between $q$
and $\bar q$ is
\begin{eqnarray}
V(r) =  {- \alpha_{\rm eff} \over r}, 
\end{eqnarray}
where $\alpha_{\rm eff} = C_f \alpha_s$, 
$\alpha_s$ is the usual QCD running coupling constant,
and
$C_f$ is 
\begin{eqnarray}
C_f = \cases{ 4/3,  & for color-singlet; \cr
             -1/6,  & for color-octet.   \cr}
\end{eqnarray}
For outgoing states relevant to a produced $q\bar q$ pair, the wave
function of the quark in the field of the antiquark is given by
\cite{Akh65}
\begin{eqnarray}
\psi = N e^{i {\hbox{\boldmath $p$}} \cdot 
{\hbox{\boldmath $r$}}}
(1- {i \over 2E} {\hbox{\boldmath $\alpha$}} \cdot \nabla )
u~~ {}_1F_1(-i \xi, 1, -i(pr + {\hbox{\boldmath $p$}} \cdot 
{\hbox{\boldmath $r$}} )).
\end{eqnarray}
Here, $u$ is a  free quark spinor, $\hbox{\boldmath
$\alpha$}$ is a Dirac matrix, ${}_1F_1$ is the confluent hypergeometrical
function, $N$ is the normalization constant
\begin{eqnarray}
|N|^2= {2 \pi \xi \over 1 - e^{-2 \pi \xi} }
\end{eqnarray}
\vskip -0.3cm\noindent
where
\vskip -0.6cm
\begin{eqnarray}
\label{eq:xi}
\xi={ \alpha_{\rm eff} \over  v}, 
\end{eqnarray}
and $v$ is the asymptotic relative velocity of the quark and the
antiquark,
\begin{eqnarray}
v= {\sqrt{s^2 - 4 s m_q^2} \over s - 2 m_q^2}. 
\end{eqnarray}
The square of the wavefunction at contact, which is a generalization
of the familiar `Gamow factor' \cite{Gam28}, gives the corrective
$K$-factor
\begin{eqnarray}
K
=|\psi(0)|^2 =  { 2 \pi \xi \over 1 - e^{-2 \pi \xi} }  
(1 + \alpha_{\rm eff}^2),
\end{eqnarray}
where the spins of both quarks have been averaged over.

 In order to obtain a generalized correction factor that is valid not
only when $q$ and $\bar q$ have low relative velocities, but also have
large relative velocities, we use the interpolation technique
suggested by Schwinger \cite{Sch73}.  In the high energy limit, the
next-to-leading order QCD correction for $e^++ e^- \rightarrow q
+\bar q$ is given by \cite{Fie89}
\begin{eqnarray}
K =1+   {  \alpha_{\rm s} \over  \pi}
\,.
\end{eqnarray}
The transition from the low-velocity corrective factor to relativistic
velocities can be accommodated with the introduction of a function
$f(v)$
\begin{eqnarray}
f(v)= {\alpha_{\rm eff} \biggl [ {1 \over v} + v \biggl ( -1 + {3
\over 4 \pi^2} \biggr ) \biggr]}. 
\end{eqnarray} 
We can thus construct an approximate correction factor for the
color-Coulomb interaction for all relative velocities as \cite{Cha95a}
\begin{eqnarray}
\label{eq:kfac}
K =
{ 2 \pi f(v) \over { 1 - \exp[-2 \pi f(v)]} }  
(1+ \alpha_{\rm eff}^2).
\end{eqnarray}

\vspace*{0.5cm}
\epsfxsize=350pt
\includegraphics{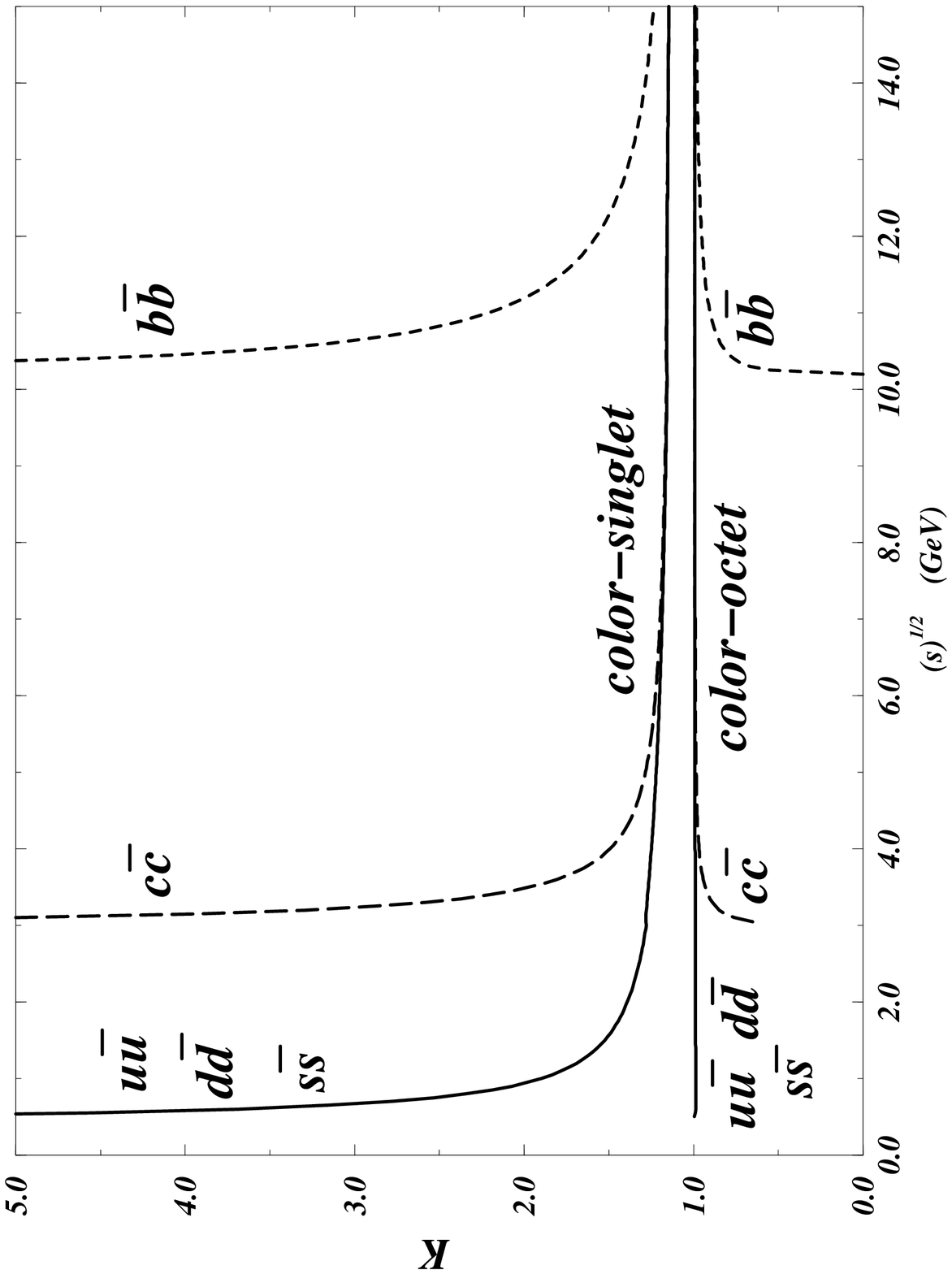}
\vskip 6.5cm
\begin{minipage}[t]{11cm}
\noindent \bf Fig.2.  \rm
The $K$-factor for $q\bar q$ production as a function of 
$\sqrt{s}$ for the color-Coulomb $q$-$\bar q$ interaction.
\end{minipage}
\vskip 4truemm

In Fig.\ 2 we show the $K$-factor calculated with Eq.\ (\ref{eq:kfac})
for the color-Coulomb interaction as a function of $\sqrt{s}$.  It
gives an enhancement for a color-singlet state and a suppression for a
color-octet state.  The correction factors deviate significantly from
unity near the threshold.  As a test of the reliability of these
correction factors, we use the $K$-factor of Eq.\ (\ref{eq:kfac}) for
the production of color-singlet $q \bar q$ pairs to calculate the
ratio $R=\sigma(e^+ e^- \rightarrow {\rm hadrons}) /\sigma( e^+ e^-
\rightarrow \mu^+ \mu^-)$ \cite{Cha95b}.  The results of $R$ ratios
are given as the solid curve in Fig.\ 1, which are in reasonable
agreement with experiment.  This agreement lends support to the
application of the $K$-factor (\ref{eq:kfac}) to other processes.

\section{K-factor for a Color-Yukawa Interaction} 

If one places a quark and an antiquark in a quark-gluon plasma, their
interaction will be screened to become a color-Yukawa interaction:
\begin{eqnarray}
\label{eq:pot}
V(r)=-{ \alpha_{\rm eff}  e^{-r/\lambda_{{}_D}} \over r} \,,
\end{eqnarray}
where the Debye screening length $\lambda_{{}_D}$ is inversely
proportional to the quark-gluon plasma temperature $T$ \cite{Gro81}.
To obtain the corrective $K$-factor we evaluate the wave function for
a quark and an antiquark in the color-Yukawa potential using the
phase-amplitude method of Calogero\cite{Cal67}.
\vspace*{1.5cm}
\epsfxsize=350pt
\includegraphics{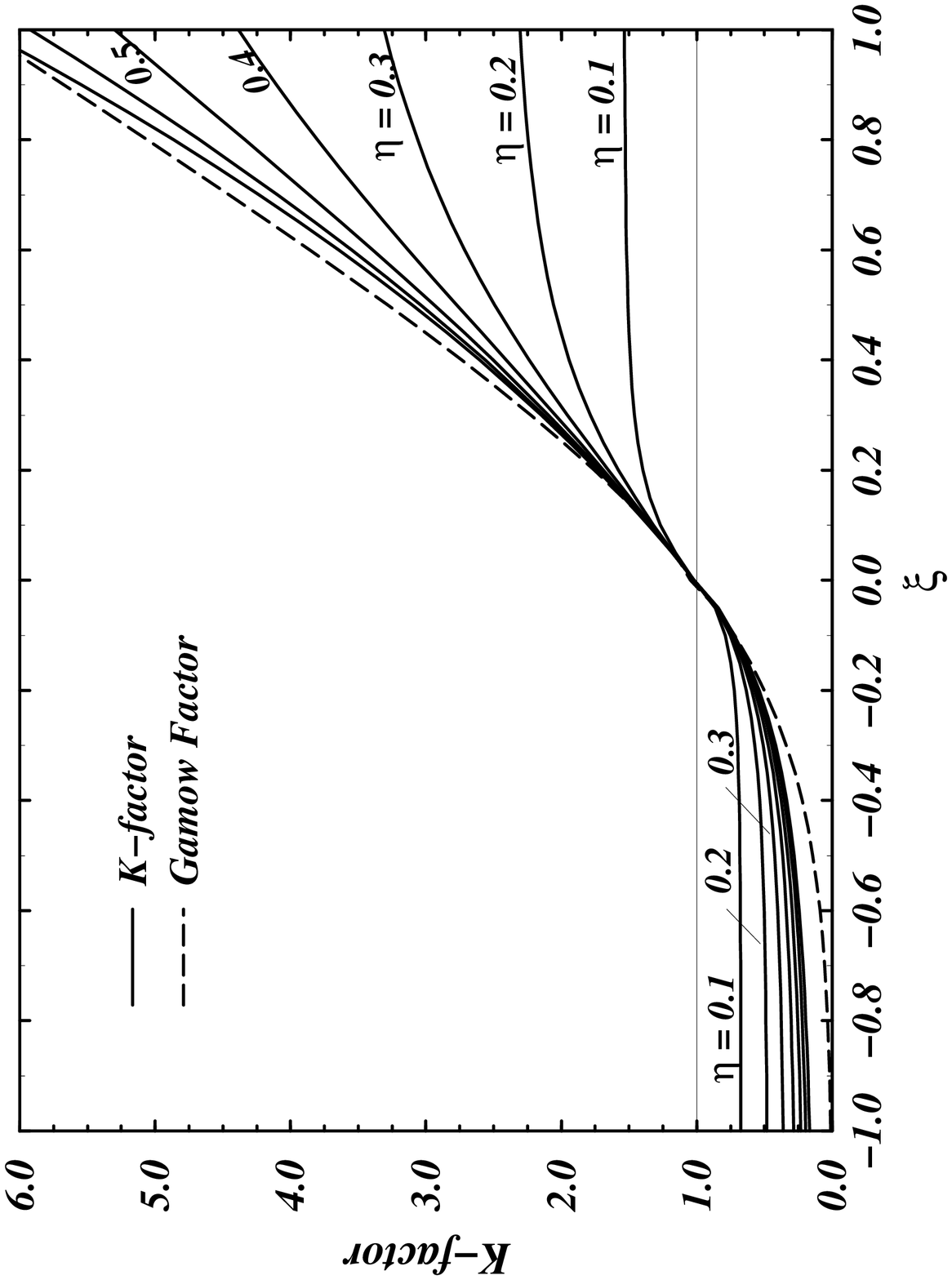}
\vskip 5.0cm
\begin{minipage}[t]{11cm}
\noindent \bf Fig.3.  \rm
The $K$-factor for $q\bar q$ production as a function of 
$\xi$ and $\eta$ for the color-Yukawa $q$-$\bar q$ interaction.
\end{minipage}
\vskip 4truemm

We introduce $a_{{}_B}=1/\mu|\alpha_{\rm eff}|$, where $\mu$ is the
reduced mass.  The quantity $a_{{}_B}$ is the Bohr radius for an
attractive potential and the appropriate length scale for a repulsive
potential.  The Schr\"odinger equation in $x=r/a_{{}_B}$ contains only
the Coulomb parameter $ \xi= {\alpha_{\rm eff} / v} $ and the
dimensionless screening length parameter
\begin{eqnarray}
\eta={ \lambda_{{}_D} \over a_{{}_B}} \,.
\end{eqnarray}
The $K$-factor is therefore only a function of $\xi$ and $\eta$.  The
quantity $K(\xi,\eta)$ has been evaluated in Ref.\ \cite{Won96c}, and
we summarize the main results here.

In Fig.\ 3 we display $K(\xi,\eta)$ for the color-Yukawa potential for
$-1 \le \xi \le 1$ and $\eta=0.1, 0.2,...,0.7$.  It is greater than
unity for positive $\xi$, corresponding to an attractive interaction.
It is less than unity for negative $\xi$, corresponding to a repulsive
interaction.  In the limit $\eta \rightarrow \infty$, the corrective
factor $K(\xi,\eta)$ approaches the Gamow factor, which is a function
of $\xi$ only.

In Fig.\ 4a we show $K(\xi,\eta)$ as a function of $\eta$ over the
range $0.4 \le \eta \le 2$.  We find that $K(\xi,\eta)$ has a maximum
at $\eta=0.835$.  The larger the values of $\xi$, the greater is the
maximum of $K(\xi,\eta)$.  The peak values of the $K$-factor are
much greater than the corresponding Gamow factor, as indicated by the
ratio $K(\xi,\eta)/$(Gamow factor) in Fig.\
4b.
\vspace*{2.5cm}
\epsfxsize=200pt
\includegraphics{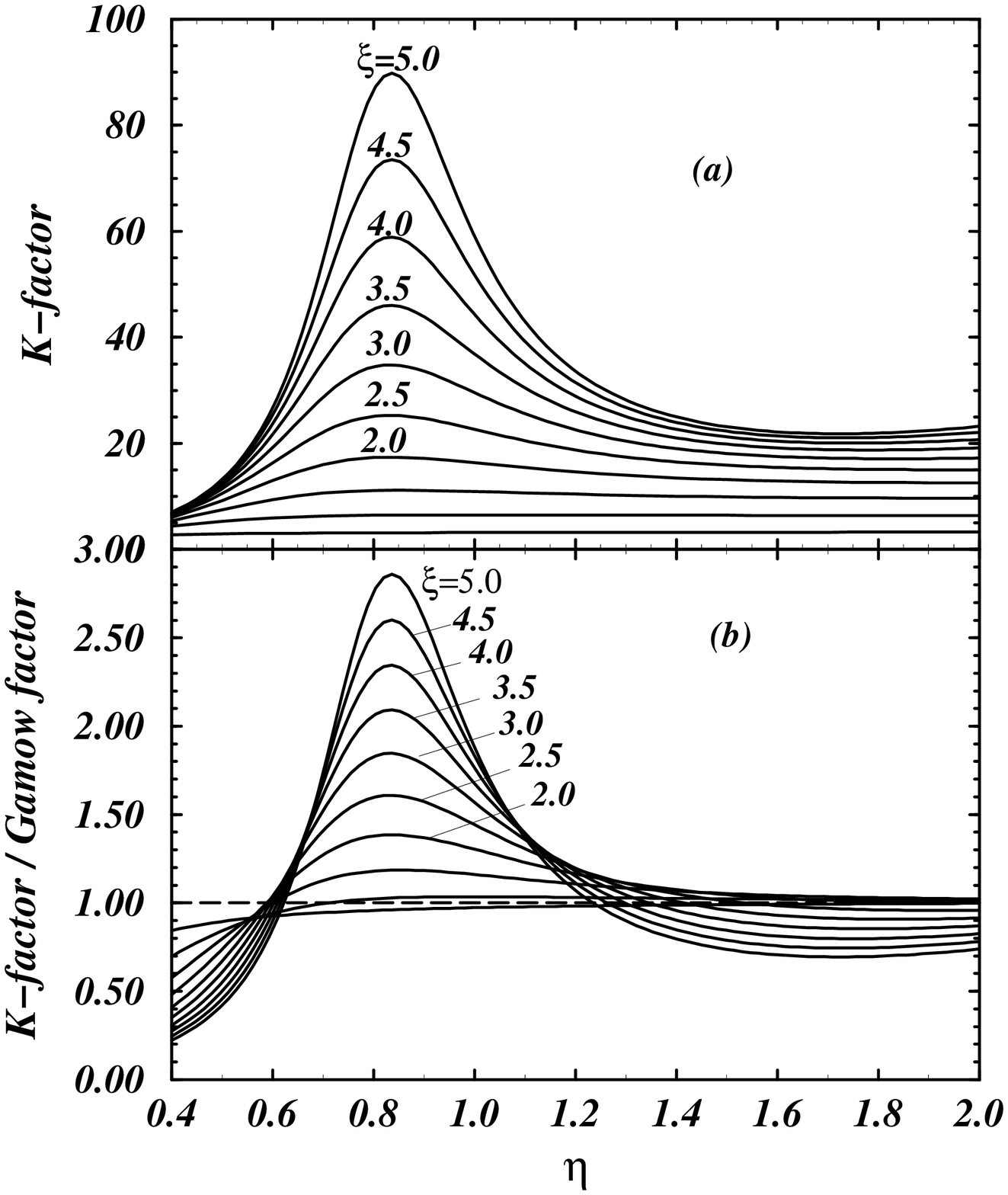}
\vskip 6.5cm
\begin{minipage}[t]{11cm}
\noindent \bf Fig.4.  \rm {\bf (a)} The $K$-factor for $q\bar q$
production as a function of $\xi$ and $\eta$ for the color-Yukawa
$q$-$\bar q$ interaction.  {\bf (b)} The ratio of $K$ to the Gamow
factor as a function of $\xi$ and $\eta$.
\end{minipage}
\vskip 4truemm

The prominent peak of $K(\xi,\eta)$ at $\eta=0.835$ is due to the
emergence of the lowest bound state of the system into the continuum
to become a $q\bar q$ resonance as the screening length
decreases. Movement of bound states into the continuum is familiar in
the context of a single-particle system in a finite-range potential,
as in nuclear single-particle states \cite{Boh67} and in the
scattering of an electron from an atom\cite{Mor32}.  Numerical
calculations show that $\eta\approx 0.84$ is the screening length
parameter for which the lowest bound state in the screened potential
becomes unbound \cite{Mat88}.  This coincides with the location of the
peaks of $K(\xi,\eta)$ in $\eta$.  We have found that there is a
similar peak of $K(\xi,\eta)$ at $\eta=3.23$, corresponding to the
second $s$-wave bound state emerging into the continuum to become a
$q\bar q$ resonance.

\section {Strange and charm $q\bar q$  production}

Basic processes for $q \bar q$ production in nucleon-nucleon
collisions and in a quark-gluon plasma are $g g \rightarrow q \bar q ~
$ and $q \bar q \rightarrow g^* \rightarrow q \bar q~ $.  We can use
the $K$-factor to correct the lowest-order result of Ref.\
\cite{Com79} for $q \bar q$ production in these processes as well. The
corrected cross section for $g g \rightarrow q\bar q$, (where $q$
refers to $s$ or $c$ quarks), averaged over initial gluon types and
polarizations and summed over final colors and spins, is
$$
\sigma_{gg} (M_{q\bar q}) = K_{gg} { \pi\alpha^2_s \over 3 M_{q\bar
q}^2}\biggl\{ (1 +\eta_m +{1\over 16} \eta_m^2) \ln
\biggl({1+\sqrt{1-\eta_m}\over 1-\sqrt{1-\eta_m}} \biggr) 
~~~~~~~~~~~~~~
$$
\vskip -0.55cm
\begin{eqnarray}
\label{gg}
~~~~~~~~~~~~~~~~~~~~~~~~~~~~~~~~~~~~~- \biggl(
{7\over 4} + {31\over16} \eta_m \biggr) \sqrt{1-\eta_m}
\biggr)\biggr\}\,,
\end{eqnarray}
where $\eta_m =4 m^2_q/M_{q\bar q}^2$, $m_q$ is the mass of the quark
and $M_{q\bar q}$ is the invariant mass of the produced $q\bar q$
pair.  For $q\bar q$ production by gluon fusion, $q\bar q$ pairs are
produced with a relative color-octet/color-singlet ratio of
\cite{Fad90}
\begin{eqnarray}
{ {\rm (color-octet)} \over {\rm (color-singlet)}}=
{(d^{\rm{abc}}/\sqrt{2})^2 \over (\delta^{\rm{ab}}/\sqrt{3})^2} ={5\over 2}\,.
\end{eqnarray}
Taking these weights into account, we can write the corrective factor
for the gluon fusion mode as $K_{gg}= [5K({\rm octet}) +2K({\rm
singlet})]/7$.

Similarly, the corrected cross section for $q\bar q \rightarrow g^*
\rightarrow q\bar q$, (where $q$ refers to $s$ or $c$ quarks),
averaged over initial and summed over final colors and spins, can be
written as
\begin{eqnarray}
\label{eq:qq}
\sigma_{q\bar q} (M_{q\bar q}) =  K_{q\bar q} 
{ 8 \pi\alpha^2_s \over 27 M_{q\bar q}^2}
\biggl (1+{\eta_m
\over 2}\biggr) \sqrt{1-\eta_m}\,.
\end{eqnarray}
For $q\bar q$ production by quark-antiquark annihilation
through a virtual gluon in lowest order, the produced $q\bar q$
pair is in a color-octet state.  The corrective factor to be used is
the color-octet corrective factor, $K_{q\bar q}= K({\rm color~octet})$.

As illustrative examples we shall study two cases with different
plasma screening lengths.  For the quark-gluon plasma with $N_f$
flavors and $N_c=3$, lowest-order perturbative QCD gives a Debye
screening length of \cite{Gro81}
\begin{eqnarray}
\lambda_{{}_D}({\rm PQCD})
=
{1 \over \sqrt{  \bigl ( {  N_c \over 3}+ {N_f \over 6 }\bigr ) g^2  }
~T} \,.
\end{eqnarray}
For a coupling constant $\alpha_s=g^2/4\pi=0.3$ and $N_f=3$, the Debye
screening length at a temperature of 200 MeV is $\lambda_{{}_D}
\approx 0.4$ fm.  We shall examine the cases of $\lambda_{{}_D}=0.2$ and 0.4
fm, corresponding respectively to temperatures of 400 MeV and 200 MeV
in this perturbative QCD estimate. 

The final-state interaction enhances the production cross section for
color-singlet states, and suppresses the production for color-octet
states. The gluon fusion modes produce $q\bar q$ pairs in a
superposition of color-singlet and color-octet states, and the larger
magnitude of the color-singlet $K$-factor dominates the correction
even though the singlet weight factor is lower.  The net result is an
enhancement of production by gluon-fusion.  On the other hand, for
$s\bar s$ and $c\bar c$ production by $q\bar q$ annihilation, the
intermediate virtual gluon selects color-octet states only. The color
interaction is repulsive, so the cross section is suppressed.

\vspace*{1.5cm}
\hskip -1.0cm
\epsfxsize=200pt
\includegraphics{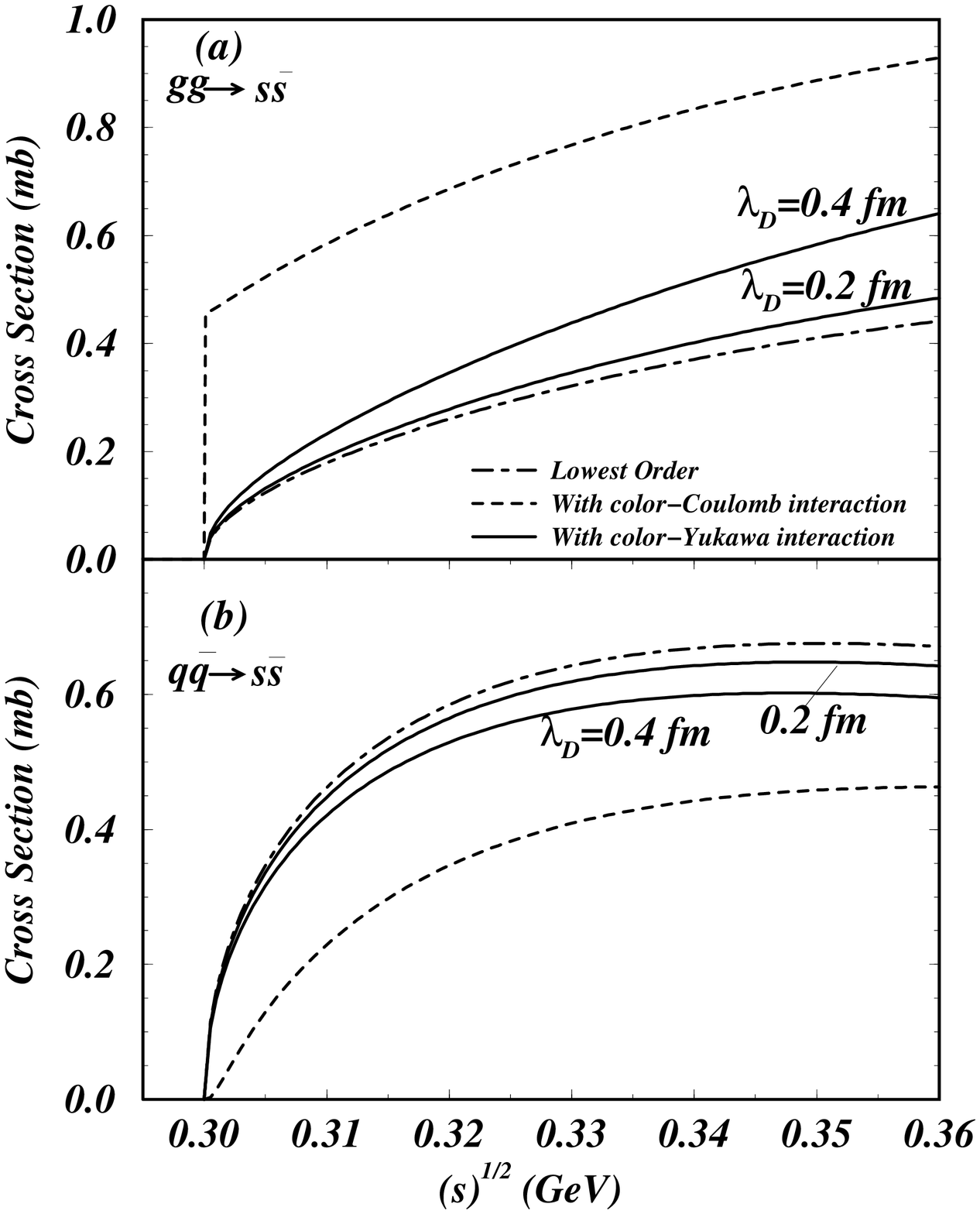}
\vskip 7.2cm
\begin{minipage}[t]{11cm}
\noindent \bf Fig.5.  \rm Cross sections for $s \bar s$ 
production {\bf (a)} by gluon fusion, and {\bf (b)} by $q\bar q$
annihilation.
\end{minipage}
\vskip 4truemm

In Figures 5 we present the cross sections for $s \bar s$ production
in a quark-gluon plasma with screening lengths of 0.2 and 0.4 fm
(solid curves).  The cross sections for the color-Coulomb interaction
without screening are shown as dashed curves.  The lowest-order cross
sections are shown as dashed-dot curves for comparison.

For an $s\bar s$ pair, $a_B$ is about 3 fm for a color-singlet state
and about 24 fm for a color-octet state.  Therefore, the relevant
screening length parameter $\eta$ is small for the cases of
$\lambda=0.2$ and 0.4 fm.  This implies that screening reduces the
final-state interaction significantly, and $K\approx 1$ for $s \bar s$
production in a quark-gluon plasma.  For a plasma with a screening
length of 0.2 to 0.4 fm, the effect of screening reduces the
final-state interaction so that the cross sections are close to the
lowest-order cross sections.  The situation is quite different from
the case without screening.  The final-state color-Coulomb interaction
leads to a $K$-factor which is significantly different from unity.  As
shown in Fig.\ 5, in the case with final-state color-Coulomb
interaction, the $s\bar s$ production cross section through gluon
fusion is greatly enhanced, while the production through $q\bar q$
annihilation is suppressed from the lowest-order cross sections.
\vspace*{1.5cm}
\hskip -1.0cm
\epsfxsize=200pt
\includegraphics{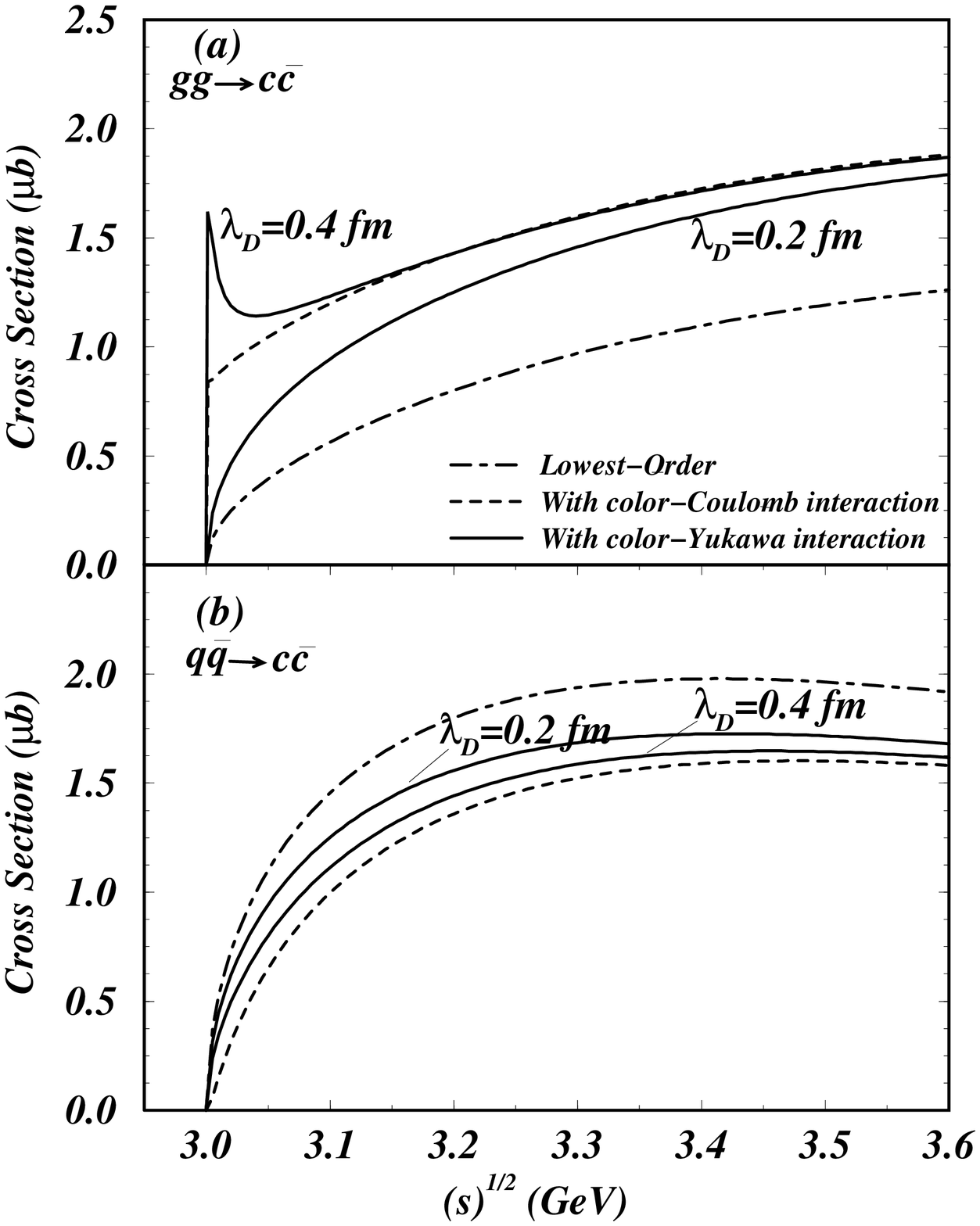}
\vskip 8.0cm
\begin{minipage}[t]{11cm}
\noindent \bf Fig.6.  \rm
Cross sections for $c \bar c$ production {\bf (a)} by gluon fusion,
and {\bf (b)} by $q\bar q$ annihilation.
\end{minipage}
\vskip 4truemm

The effect of screening manifests itself for $c \bar c$ production in
a different way, as shown in Fig.\ 6.  The quantity $a_B$ for a $c\bar
c$ pair is about 0.5 fm for a color-singlet state, and about 4 fm for
a color-octet state.  Thus, for a color-singlet $c\bar c$ system, the
screening length parameter for $\lambda=0.4$ fm is $\eta\approx 0.8$,
close to the value $\eta=0.834$ for the occurrence of the production
resonance.  Thus, when the screening length is near 0.4 fm, screening
greatly enhances the production cross section for color-singlet $c\bar
c$ near threshold.  The theoretical $c\bar c$ production cross
sections through the gluon fusion mode therefore exhibit a marked rise
at small velocities (Fig.\ 6a).

For $c\bar c$ production by $q\bar q$ annihilation, the intermediate
virtual gluon selects color-octet states only.  The color interaction
is therefore repulsive, and suppresses the cross section.  With
typical screening lengths of 0.2 and 0.4 fm, the corrected cross
sections are near the color-Coulomb values for $c \bar c$, while for
$s \bar s$ production they are close to the lowest-order tree-level
values, in accordance with the different values of $\eta$ for the two
flavors.

\section{Conclusions and Discussions}

In reactions involving $q\bar q$ production, the quark and the
antiquark are subject to final-state interactions due to their mutual
interaction.  The lowest-order cross sections for these processes can
be corrected by using an approximate corrective $K$-factor to take
into account the $q$-$\bar q$ interaction.  We have obtained the
$K$-factor for final-state color-Coulomb and color-Yukawa
interactions.  The corrective $K$-factor for the color-Yukawa
interaction depends on two dimensionless parameters: the usual Coulomb
parameter $\xi=\alpha_{\rm eff}/v$, and the screening length parameter
$\eta=\lambda_{{}_D}/a_{{}_B}$.  For attractive Yukawa potentials we
observe prominent peaks of the $K$-factor as a function of the
screening length parameter $\eta$.  The peaks are located at
$\eta=0.835$ and $\eta=3.23$, corresponding to two lowest $s$-wave $q
\bar q$ bound states emerging into the continuum to become $q\bar q$
resonances as the screening length decreases.  We have calculated the
cross sections for two typical choices of the Debye screening length
($0.2$ fm and $0.4$ fm), corresponding to plasma temperatures of
approximately $400$ and $200$ MeV respectively.  While the corrections
to the cross section in the color-Coulomb limit are of similar
magnitude for $s\bar s$ and $c\bar c$ pairs in the same velocity
range, they are considerably different for the two systems in the
presence of screening with a color-Yukawa interaction.  This arises
because the quantity $a_B$ is much smaller for the $c$-$\bar c$ system
than for the $s$-$\bar s$ system.  For the color-singlet case, a
screening length of 0.4 fm corresponds to $\eta\approx 0.8$, which is
near a zero-energy $c$-$\bar c$ resonance.  This is in contrast to $s
\bar s$ production for $\lambda_{{}_D}=0.2-0.4$ fm, for which the
corresponding screening length parameters of $\eta=0.04-0.08$ are very
small, and are far from $\eta=0.835$ for a $q\bar q$ resonance.

High-energy heavy-ion collisions have become the focus of intense
research because of the possibility of producing a quark-gluon plasma
during such collisions. The suppression of $J/\psi$ production has
been suggested as a probe of the screening between a charm quark and a
charm antiquark in the plasma, because $J/\psi$ production is
suppressed in a quark-gluon plasma above a temperature $T_{c \bar c}$,
such that $c$ and $\bar c$ cannot form a bound state
\cite{Mat86,Mat88}.  The presence of the quark-gluon plasma is
signalled by a substantial decrease of the probability for $J/\psi$
production above $T_{c\bar c}$.

From our results, the occurrence of $q\bar q$ resonances may provide a
complementary signal to search for the quark-gluon plasma.  These
$q\bar q$ resonances give rise to prominent peaks of the $K$-factor as
a function of the screening length parameter, which is the ratio of
the screening length $\lambda_{{}_D}$ to $a_{{}_B}$.  The screening
length is inversely proportional to the temperature \cite{Gro81}.
Thus, $q\bar q$ resonances lead to prominent peaks of the $K$-factor
at specific plasma temperatures.  We have seen in Fig.\ 6 that large
values of the $K$-factor near the threshold give rise to a narrow peak
in the heavy-quark production cross section just above the threshold.
The occurrence of a $q \bar q$ resonance will be accompanied by a much
enhanced $q\bar q$ production cross section just above the
threshold. The enhancement will be a function of the temperature.
Here, the quark-gluon plasma is signalled by the presence of a $c\bar
c$ resonance just above the threshold at $T_{c\bar c}$.

The search for $q\bar q$ screening resonances in the quark-gluon
plasma can make use of the peaks of the $K$-factor at $\eta=0.835$ and
$\eta=$3.23.  The resonance at $\eta=3.23$ may not lead to realizable
enhancements because it corresponds to temperatures much below the
estimated quark-gluon plasma transition temperature (of approximately
$150-200$ MeV).  Using the perturbative QCD estimates, the screening
length parameter $\eta=0.835$ corresponds to a $c$-$\bar c$
resonance at $T_{c\bar c} \approx 165$ MeV and a $b \bar b$ resonance
at $T_{b\bar b}\approx 393$ MeV.  These $T_{c\bar c}$ and $T_{b\bar
b}$ estimates from PQCD are approximate and uncertain, as lattice
gauge theory gives Debye screening lengths of about half the PQCD
estimates \cite{Uka89}.  The Debye screening length needs to be
determined by experimental searches for these $c\bar c$ and $b\bar b$
resonances using the peaks in the $K$-factors.  Temperature dependence
of this type arises from the nature of screening between the
interacting heavy quark and its antiquark partner, which is an
important property to identify the deconfined quark-gluon plasma.  A
search for the quark-gluon plasma using heavy-quark resonances will
require the measurement of the production yield of heavy-quark pairs
near the threshold, and a method to estimate the temperature of the
environment in which the production takes place.  The enhancement will
occur either for the production of heavy-quark pairs by the collision
of the constituents of the thermalized quark-gluon plasma, or by the
collision of the partons in nucleon-nucleon collisions in a
quark-gluon plasma environment.

\vskip 10pt
\noindent \large {\bf Acknowledgement}\normalsize
\vskip 10pt
\noindent
Lali Chatterjee would like to thank Dr. F. Plasil and Dr. M. Strayer
for their kind hospitality at ORNL.  This research was supported by
the Division of Nuclear Physics, U.S. D. O. E.  under Contract
DE-AC05-96OR22464 managed by Lockheed Martin Energy Research Corp.

\vfill\eject 

\newpage

\begin{thebibliography}{99}\parindent=8truemm
\itemsep -1mm

\bibitem{Raf96} J. Rafelski, J. Letessier, and A. Tounsi, Acta Physica
Polonica {\bf B27}, 1037 (1996).

\bibitem{Won94} C. Y. Wong, {\it
    Introduction to High-Energy Heavy-Ion Collisions}, World
  Scientific Publishing Company, 1994.


\bibitem{Mos95} M. G.-H. Mostafa, C. Y. Wong, L.  Chatterjee, 
 and Zhong-Qi Wang, ORNL Preprint ORNL-CTP-95-03/hep-ph-9503357, 
to be published in International Journal of Modern Physics.

\bibitem{Gav96}
S. Gavin, P. L. McGaughey, P. V. Ruuskanen, and R. Vogt,
LBL Preprint 37981, hep-ph/9604369.


\bibitem{Fie89} R.\ D.\ Field, {\it Applications of Perturbative
    QCD}, Addison-Wesley Publishing Company, 1989.

\bibitem{Bar80} R. M. Barnett, M. Dine, and L. McLerran,{\it
Phys. Rev.} {\bf D22} (1980) 594.

\bibitem{Gus88} S. G\"usken, J. H. K\"uhn, and P. M. Zerwas, {\it
Phys. Lett.} {\bf B155} (1988) 185.

\bibitem{Fad88}
V. Fadin and V. Khoze, {\it Sov. Jour. Nucl. Phys.} {\bf 48} (1988) 487.


\bibitem{Fad90} V. Fadin, V. Khoze, and T. Sj\"ostrand, {\it
Zeit. Phys.} {\bf C48} (1990) 613.


\bibitem{Cha95a} L. Chatterjee and C. Y. Wong, {\it Phys. Rev.} {\bf
 C51} (1995) 2125.

\bibitem{Cha95b} L. Chatterjee and C. Y. Wong, hep-ph/9501218. 

 
\bibitem{Won96c} C. Y. Wong and  L. Chatterjee, ORNL Preprint
ORNL-CTP-96-01, hep-ph/9604224.

\bibitem{Gam28}
G. Gamow, {\it Zeit. Phys.} {\bf 51} (1928) 204, see also 
L. I. Schiff, {\it Quantum Mechanics}, McGraw-Hill Company, 1955, p. 142.
A. Sommerfeld, {\it Atmobau und Spektralinien}, Bd. 2. Braunschweig:
Vieweg 1939.

\bibitem{Gyu81} M. Gyulassy and S. K. Kaufmann, {\it Nucl. Phys.} {\bf
A362} (1981) 503.

\bibitem{Akh65}
A. I. Akhiezer and V. B. Berestetskii, {\it  Quantum Electrodynamics},
Interscience Publishers, New York, 1965.

\bibitem{Sch73} J. Schwinger, {\it Particles, Sources, and Fields
    } (Addison-Wesley, New York, 1973), Vol. II, Chaps. 4 and 5.


\bibitem{Gro81} D. Gross, R. D. Pisarski, and L. G. Yaffe, {\it
Rev. Mod. Phys.} {\bf 53} (1981) 43.

\bibitem{Cal67} F. Calogero, {\it Variable Phase Approach to Potential
Scattering},  Academic Press, N.Y. 1967.

\bibitem{Mor32} P. M. Morse, {\it Rev. Mod. Phys.} {\bf 4} (1932) 577;
 W. P. Allis and P. M. Morse, {\it Zeit. Phys.}  {\bf 70} (1931) 567.

\bibitem{Mat88} T. Matsui, {\it Zeit. Phys.} {\bf C38} (1988) 245.

\bibitem{Boh67}
A. Bohr, and B. Mottelson, {\it Nuclear Structure}, W. A. Benjamin,
Inc., New York, 1967.

\bibitem{Com79} B. L. Combridge, {\it Nucl. Phys.} {\bf B151} (1979)
429.

\bibitem{Mat86} T. Matsui and H. Satz, {\it Phys. Lett.}  {\bf B178}
(1986) 416.

\bibitem{Uka89}
A. Ukawa, {\it Nucl. Phys.} {\bf A498}  (1989) 227c.


\end{thebibliography}
\end{document}